\documentclass[12pt]{article}
\usepackage{amsmath}
\usepackage{amsfonts}
\usepackage{amssymb}
\usepackage{amscd}
\usepackage{epsfig}
\newtheorem{thm}{Theorem}
\newtheorem{stm}{Statement}
\begin{document}
\begin{center}
{\bf  Cosmological anomalies and exotic smoothness structures}\par
\vskip 1cm
Torsten A\ss elmeyer-Maluga
\footnote{e-mail: torsten@first.gmd.de}
\vskip .3cm
{\em Fraunhofer FIRST ,\\
Kekul\'estra\ss e 7,\\
12489 Berlin, Germany \\ }
\vskip .5cm
Carl H. Brans
\footnote{e-mail: brans@loyno.edu}
\vskip .3cm
{\it
Loyola University, \\
New Orleans, LA 70118, USA \\
}
\end{center}
\vskip 1cm
\begin{abstract}
It seems to be generally accepted that apparently anomalous cosmological observations, such as
accelerating expansion, etc., necessarily are inconsistent with
 standard general relativity and standard matter sources. Following
 the suggestions of S\l adkowski, we point out that in addition to exotic theories
 and exotic matter there is another possibility. We refer to
 exotic differential structures on ${\mathbb R}^4$ which could
 be the source of the observed anomalies without changing the Einstein
 equations or introducing strange forms of matter.
\par
\end{abstract}
\vskip 1cm
Recent cosmological observations have been interpreted as indicating an accelerating expansion of the
universe \cite{Riess:1998cb}, \cite{Perlmutter:1999rr}. This
conclusion in turn  has led to speculations that either
or both of the following must be considered:
\begin{enumerate}
\item Einstein's equations in their original, purely metric,  form in four dimensions
  are incorrect, or,
\item the matter tensor contains ``exotic'' sources, such as dark
  energy, quintessence, etc.
\end{enumerate}
The recent literature is replete with speculations along these
lines. We cite just a few representatives:
\cite{Fischler:2001yj}, \cite{Bean:2000zm}, \cite{Barrow:2000nc},
\cite{Mannheim:2001kk}, \cite{Liddle:2000dt}, \cite{Banerjee:2000mj}, etc.
 \par
In this paper we want to  revisit the suggestion of S\l adkowski\cite{Sladkowski:1999mm} that there is
another possibility based on recent mathematical discoveries in
differential topology. Specifically,
\begin{enumerate}
\setcounter{enumi}{2}
\item the coordinates $(t,r)$ of observational cosmology may not be
  smoothly extendible indefinitely into the past.
\end{enumerate}
If 3. is valid then the standard extrapolation of earth based
observations to distant phenomena may not be justified.\par
The reason for raising the conjecture 3. lies in the  discoveries in
differential topology of
 the existence of non-standard, or
exotic, global smoothness (differential) structures on topologically
trivial spaces such as ${\mathbb R}^4,$ or ${\mathbb R}^1\times S^3.$ For reviews of the subject
aimed at the physics audience see \cite{Brans:1994hq},
\cite{Brans:1994wv}, \cite{Brans:1996}.
   Let us begin here by simply
stating a strikingly counter intuitive fact as well established
mathematically:
\begin{thm}
There exist global smoothness structures on topological ${\mathbb R}^4$ which are
not diffeomorphic to the standard one. We label such manifolds
${\mathbb R}_\Theta^4.$
\end{thm}
Thus, we can label points ${\mathbb R}_\Theta^4$ with global {\em
  topological} coordinates, $(t,x,y,z)$. However, to do calculus we
need a {\em differential (smooth) structure} defined by a family of
coordinate patches with smooth coordinate transformations in their
overlap. Such a family of coordinate patches defines a smooth
structure.  One obvious one on ${\mathbb R}^4$ is the ``standard''
one defined by one coordinate patch with smooth coordinates identical to
the global topological ones. It has been known for
some time that {\em any} smooth structure is diffeomorphic
(equivalent) to the standard one for all ${\mathbb R}^n,\ n\ne 4.$
However, the conjecture that this would also be true for  physically
critical case of $n=4$ remained unsettled until the 1980's when
pioneering work by Donaldson, Freedman, Gompf, et al., established
Theorem 1.
 Thus the statement of the theorem is
that not all of $(t,x,y,z)$ can be global smooth functions in terms of
this exotic structure.\par
For our purposes a remarkable feature of these exotic  ${\mathbb
  R}_\Theta^4$'s is that each of them contains a compact set that
cannot be contained in the interior of any smoothly embedded $S^3.$
See, for example, the discussion in pages 366ff of Gompf and
Stipsicz,\cite{GomSti:1999}
\begin{thm}
For some ${\mathbb R}_\Theta^4,$ there exist global topological
coordinates $(t,x,y,z)$ and  numbers $R_1< R_2$ such that the spheres, $S_{R_0},$
defined by
$t^2+x^2+y^2+z^2=R_0^2$ are smooth
 for $R_0<R_1,$ but are not smooth\footnote{By ``not smooth'' we mean ``not smoothly embedded.''} for
any $R_0\ge R_2.$ Choose one such, say $M,$ for our spacetime model.
\end{thm}
\par
 We can thus state that for $M$
\begin{thm}
We can choose two sets, $a$ and $b$ in $M$
  such that both cannot be  included in one smooth coordinate patch in {\em any}
  diffeomorphic presentation of $M$
\end{thm}\par
Now  look at null
geodesics between points in these two sets (see figure)  and attempts to interpret
information received in $a$ from $b$   in
terms of the a priori assumption that one coordinate patch including
the pair exists.     \par
\begin{figure}
\begin{center}
\epsfig{figure=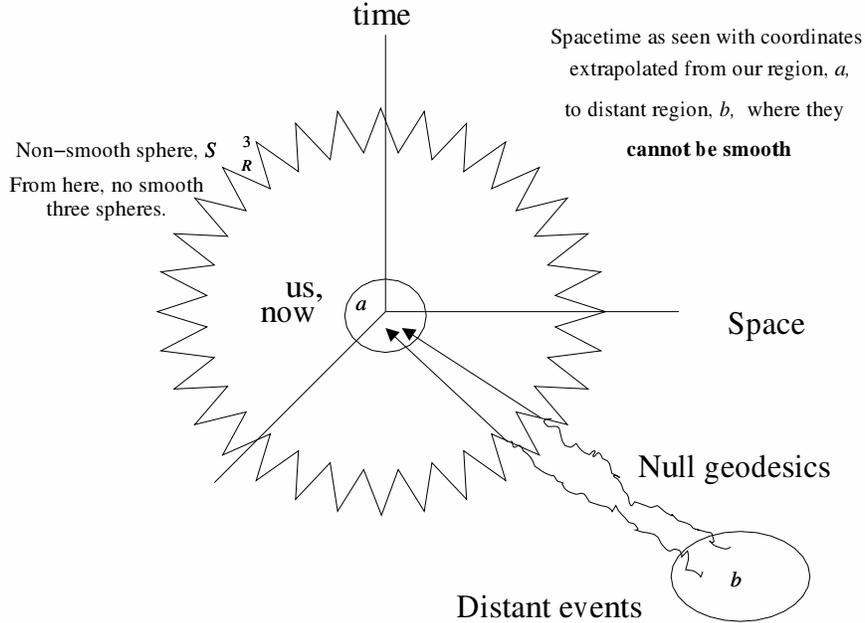,width=13cm}
\caption{Exotic spacetime with non-smooth $S^3_{R_2}$. Coordinates cannot
be smoothly continued past limiting topological $S_{R_2}^3$.}
\end{center}
\end{figure}
In looking at the figure, we must point out that the light rays, null
geodesics, are smooth {\em locally}. The reason they are represented
as non-smooth in this drawing is to point out that an astronomer trying to draw such a
figure based on his observations in $a$ alone would be
forced to use non-smooth lines since the local smooth coordinates in
$a$ can {\em not} be smoothly continued to $b$. In
other words, no smooth image such as in this figure can be drawn for
our $M={\mathbb R}^4_\Theta.$  The null
geodesics from $b$ will still be smooth and well behaved throughout
their length, and the Einstein equations satisfied with normal matter.
However, it will be incorrect to assume that we can
extrapolate from these incoming geodesics in $a$   information about $b$
{\em because we do not know the non-trivial transition function
  between the smooth coordinate patches linking the two sets}.\par

More specifically, in
observational astronomy it is generally assumed that the metric can be
written in the FRW form
\begin{equation}
ds^2=-dt^2+a(t)^2d\sigma_3^2,\label{frw1}\end{equation}
where the spatial three metric is usually expressed in spherical
coordinates in a form depending on assumptions of isotropy and
homogeneity. The associated topology is thus ${\mathbb R}^1\times M^3$
for some three-manifold, $M^3.$  In the standard models the three
metric is one of the three constant curvature ones, each containing a
``radial'' coordinate\footnote{Of course in the spherical case
  the ``radial'' coordinate is not indefinitely continuable because it
  is essentially an angular one. However, this is not the sort of
  coordinate anomaly we are addressing here and can certainly be
  accommodated in standard models.}, $r$. Because of isotropy, the incoming
geodesics are described globally (modulo the proviso in the footnote)
by differential equations involving $r,t$ only.  However, if $M$ is as
described in Theorem 3 and the figure, these may not be globally
smooth. Hence the actual metric would have to be expressed in terms of
more than one $r,t$ coordinate region, and information extracted from
the coordinate overlaps. Unfortunately, because the present
mathematical technology does not provide us with an effective
coordinate patch structure, more explicit statements than this cannot
now be made. Nevertheless,
  the assumption that we can extrapolate
information coming from incoming light rays back in time and out in
space as if these geodesics would act as a radial type of coordinate
system when indefinitely extended into their past is not valid if $M$
is used as a spacetime model. We should also note that although we
have discussed only the ${\mathbb R}_\Theta^4$ (which is actually
${\mathbb R}^1\times_{\Theta}{\mathbb R}^3$) we could equally have
chosen an exotic ${\mathbb R}^1\times_{\Theta}S^3$. \par
A simple analogy is provided by gravitational lensing phenomena. Here
we see two incoming null geodesics arriving at earth from different
directions. However, the possibility that in some reasonable situations they cannot be extrapolated
backward as ``good'' radial coordinates because they have been
focused by the gravitational lens effect of an intervening massive
object has been widely discussed and generally accepted as
viable. Thus the extrapolation of the different angle data for the two
incoming geodesics to different sources is incorrect. \par
\begin{stm}
{\bf (Gravitational lensing analogy)} Null geodesics arriving from different angles may intersect in the past
  because of gravitational curvature  caused by intervening mass
 and thus may not be extrapolated back as good radial coordinate lines.
  \end{stm}

What we are proposing
here is more radical, of course, but just as viable in the sense that
we know of no physical principles to exclude it, and it could lead to
an understanding of apparent anomalous distant time behavior without
introducing exotic theories or matter, just exotic smoothness of the
spacetime manifold model.
\par
\begin{stm}
{\bf (Exotic structures)} Null geodesics arriving from distant sources    may
 not be extrapolated back as good radial coordinate lines because of
 intervening coordinate patch transformations caused by global exotic
 smoothness.
 \end{stm}\par
 In summary, what we want to emphasize is that without changing the
 Einstein equations or introducing exotic, yet undiscovered forms of
 matter, or even without changing topology, there is a vast resource
 of possible explanations for  recently observed
surprising astrophysical data at the cosmological scale provided by differential topology.\par
While it is true that at this stage of development of the mathematical technology
 it is not possible to give explicitly the coordinate patch overlap
 functions, research along these lines is being actively
 pursued. Furthermore, S\l adkowski \cite{Sladkowski:2001}, has shown
 that it is possible to relate isometry groups (geometry) to
 differential structures in some cases. 
 \par

\end{document}